\newcommand{\eps}{\epsilon}
\newcommand{\epsl}{\eps_{l_1' + l' - l_2'}}
\newcommand{\epsli}{\eps_{l' - l_2'}}

\newcommand{\lvb}{\Bigl \bracevert}

\documentclass[pre, amssymb, showkeys, showpacs]{revtex4}
\usepackage{graphicx}
\begin{document}

\title{The Error and Repair Catastrophes:  A Two-Dimensional Phase Diagram
in the Quasispecies Model}

\author{Emmanuel Tannenbaum}
\email{etannenb@fas.harvard.edu}
\author{Eugene I. Shakhnovich}
\affiliation{Harvard University, Cambridge, MA 02138}

\begin{abstract}

This paper develops a two gene, single fitness peak model for
determining the equilibrium distribution of genotypes in a
unicellular population which is capable of genetic damage repair.
The first gene, denoted by $ \sigma_{via} $, yields a viable organism with
first order growth rate constant $ k > 1 $ if it is equal to some
target ``master'' sequence $ \sigma_{via, 0} $. The second gene, 
denoted by $ \sigma_{rep} $, yields an organism capable of genetic repair if
it is equal to some target ``master'' sequence $ \sigma_{rep, 0} $.
This model is analytically solvable in the limit of infinite
sequence length, and gives an equilibrium distribution which
depends on $ \mu \equiv L\eps $, the product of sequence length and per base
pair replication error probability, and $ \eps_r $, the
probability of repair failure per base pair.  The equilibrium
distribution is shown to exist in one of three possible ``phases.''
In the first phase, the population is localized about the 
viability and repairing master sequences.  As $ \eps_r $ exceeds the fraction 
of deleterious mutations, the population undergoes a ``repair''
catastrophe, in which the equilibrium distribution is still
localized about the viability master sequence, but is spread
ergodically over the sequence subspace defined by the repair gene.
Below the repair catastrophe, the distribution undergoes the
error catastrophe when $ \mu $ exceeds $ \ln k/\eps_r $, while
above the repair catastrophe, the distribution undergoes the error
catastrophe when $ \mu $ exceeds $ \ln k/f_{del} $, where $
f_{del} $ denotes the fraction of deleterious mutations.

\end{abstract}



\maketitle

\section{Introduction}

To cope with genetic damage to their genomes, cellular organisms
have developed a host of mechanisms to repair, and, if necessary,
replace, damaged DNA.  Environmental damage due to mutagens,
metabolic free radicals, and radiation is repaired by enzymes
which continuously scan the DNA molecule and repair the damaged
portions.  Replication errors are also repaired by several
methods.  In {\it Escherichia coli}, the DNA replicase Pol III has
a built-in proofreading mechanism which reduces the replication
error probability to $ 10^{-7} - 10^{-6} $ per base pair.
Furthermore, immediately following replication, a second
proofreading mechanism, known as mismatch repair, identifies and
corrects mismatched base pairs. In {\it E. coli}, the mismatch
repair system reduces the error probability to $ 10^{-10} -
10^{-8} $ per base pair \cite{VOET}.

The DNA mismatch repair system is of considerable interest because
it is believed that mismatch repair deficient strains, or
mutators, play an important role in the emergence of antibiotic drug
resistance and cancer \cite{MUT1, MUT2, MUT3, MUT4, MUT5, MUT6}.  Because 
mutators have mutation rates which are $ 10 $ to $ 10,000 $ times higher than 
wild-type strains, they can more rapidly adapt to hostile environments, 
thereby explaining their potential importance in understanding drug resistance.
However, mutators can accumulate genetic damage much more rapidly
than nonmutators, and hence can serve as an intermediate for the
appearance of cancerous cells in multicellular organisms.

In an earlier work, we developed a simple, analytically solvable
model to determine the equilibrium population of mutators in an
asexual, unicellular population of replicating organisms \cite{TDS}.  
The main result of this model was that at equilibrium, the population
can exist in one of two ``phases.''  For sufficiently efficient repair, 
the population was shown to exist in a ``repairer'' phase, in which the
fraction of repairers is a finite, positive quantity which depends only
on the efficiency of repair and the fraction of the genome coding for
repair.  The equilibrium genotype of the population is localized about
the ``master'' subsequence for which repair is functionining.  When the
repair efficiency drops below a critical value, the population delocalizes
over the repair sequence subspace, and the fraction of repairers becomes zero 
in the limit of infinite genome length.  This phase is naturally termed 
the ``mutator'' phase.  In \cite{TDS} the transition from the repairer
to the mutator phases was called the repair catastrophe.

The solution of the model presented in \cite{TDS} is incomplete, in
that it describes the equilibrium behavior of the system in the
low mutation rate regime.  This allowed one to assume that only
point mutations were important, considerably simplifying the
calculations.  However, another phase transition has also been
shown to occur when the mutation rate becomes too large.  Above a
critical mutation rate, replicative selection can no longer
recover the loss of information due to genetic damage.
This phenomenon is known as the error catastrophe, and was first
predicted to occur by Eigen \cite{EIG1, EIG2}.  It has since been studied in a
number of theoretical papers \cite{SFP1, SFP2, SFP3, QUAS1, QUAS2, QUAS3} 
(and references therein), and also observed experimentally \cite{CAT1, CAT2}.  

Because the model presented in our paper was only solved in the
point-mutation regime, it did not incorporate the effect of the
error catastrophe.  The assumption underlying our initial approach
was that mutators, despite their higher than wild-type mutation
rates, are still viable organisms, and so live well below the
error catastrophe.

The method used in our paper has since been generalized, so that it is
no longer necessary to assume only point mutations.  Therefore it is 
possible to obtain the equilibrium behavior for arbitrary mutation rates.  
Thus, the interplay between the error and repair catastrophes can be studied. 
We believe that our approach is quite powerful, and may be applied
toward solving a large class of mutation dynamics equations.

This paper is organized as follows:  In the following section, we
present a brief review of the quasispecies equations developed by
Eigen, which are often the starting point for studies in
evolutionary dynamics.  We continue in Section III by developing
the form of the quasispecies equations for our mutator model.  We solve 
this model in Section IV.  Specifically, we solve
for the equilibrium fraction of viable organisms and viable
repairers, which will allow the construction of a two dimensional
phase diagram incorporating both the error and repair
catastrophes.  We develop a recursive formula for computing the
equilibrium fraction of organisms with a given genome, and we also
study the localization of the distribution.  We also look at
limiting forms of the distribution, and compare the results in
certain cases with the corresponding results obtained in \cite{TDS}.  
Finally, in Section V we present our conclusions, and plans for 
future research.

\section{The Quasispecies Equations}

The quasispecies equations are possibly the simplest for modelling
the evolutionary dynamics of a unicellular, asexual population of
replicating organisms.  We let $ n_{\sigma} $ denote the number of
organisms with genome $ \sigma $, and $ \kappa_\sigma $ denote the
first-order growth rate constant of an organism with genome $
\sigma $.  If $ \kappa_m(\sigma, \sigma') $ is taken to be the
first-order mutation rate constant from $ \sigma $ to $ \sigma' $,
then the time evolution of $ n_{\sigma} $ is given by,
\begin{equation}
\frac{d n_{\sigma}}{dt} = \kappa_{\sigma} n_\sigma + \sum_{\sigma'
\neq \sigma}{[\kappa_m(\sigma', \sigma) n_{\sigma'} -
\kappa_m(\sigma, \sigma') n_{\sigma}]}
\end{equation}
The mapping $ K : \{\sigma\} \rightarrow \{\kappa_{\sigma} \} $
defines what is called the fitness landscape.  In general, the
fitness landscape will be time dependent, since organisms usually
live in dynamic environments.  However, since in this paper we wish 
to study equilibrium behaviors, we take the fitness landscape to be static.

The conversion to Eigen's quasispecies equations 
is accomplished by converting from absolute populations to population 
fractions.  Thus, we define $ n = \sum_{\sigma}{n_{\sigma}} $, and $
x_{\sigma} = n_{\sigma}/n $. When reexpressed in terms of the $
x_{\sigma} $, the dynamical equations become,
\begin{equation}
\frac{d x_{\sigma}}{dt} = (\kappa_{\sigma} - \bar{\kappa}(t))
x_{\sigma} + \sum_{\sigma' \neq \sigma}{[\kappa_m(\sigma', \sigma)
x_{\sigma'} - \kappa_m(\sigma, \sigma') x_{\sigma}]}
\end{equation}
where $ \bar{\kappa}(t) \equiv \sum_{\sigma}{\kappa_{\sigma}
x_{\sigma}} $.  Note then that $ \bar{\kappa}(t) $ is simply the
mean fitness of the population, and arises as a normalization term
which ensures that $ \sum_{\sigma}{x_{\sigma}} = 1 $ at all times.

If we define $ \kappa_m(\sigma, \sigma) = \kappa_{\sigma} -
\sum_{\sigma' \neq \sigma}{\kappa_m(\sigma, \sigma')} $, then the
quasispecies equations become,
\begin{equation}
\frac{d x_{\sigma}}{dt} = \sum_{\sigma'}{\kappa_m(\sigma', \sigma)
x_{\sigma'}} - \bar{\kappa}(t) x_{\sigma}
\end{equation}

We may simplify the notation further by defining $ \vec{x} =
(x_{\sigma}) $ to be the vector of population fractions, and $ A
\equiv (A_{\sigma \sigma'} = \kappa_m(\sigma', \sigma)) $ to be
the matrix of mutation rate constants.  We may also define $
\vec{\kappa} $ to be the vector of growth rate constants, so that
$ \bar{\kappa}(t) = \vec{\kappa} \cdot \vec{x} $.  Then we obtain,
\begin{equation}
\frac{d \vec{x}}{dt} = A \vec{x} - (\vec{\kappa} \cdot \vec{x})
\vec{x}
\end{equation}
Eigen showed that the system evolves to an equilibrium
distribution given by the eigenvector corresponding to the largest
eigenvalue of $ A $ \cite{EIG1, EIG2}.  If the equilibrium distribution 
is denoted by $ \vec{x}_{equil} $, and the largest eigenvalue is denoted by $
\lambda $, then it is possible to show that $ \lambda = \vec{\kappa} 
\cdot \vec{x}_{equil} $.  

To obtain an expression for $ \kappa_m(\sigma', \sigma) $, let us
assume that mutations occur due to replication errors.  We take a
per base pair replication error probability of $ \eps_{\sigma'} $
for $ \sigma' $.  Let $ l \equiv HD(\sigma, \sigma') $ denote the
Hamming distance between $ \sigma $ and $ \sigma' $.  Then
incorrect replication must occur at exactly $ l $ sites along the
$ \sigma' $ sequence, and correct replication must occur at the
remaining $ L - l $ sites, where $ L $ denotes the gene sequence
length.  For an alphabet size of $ S $, incorrect replication can
be made to one of $ S - 1 $ remaining bases.  Thus, the per base
pair probability of replication to the corresponding base pair in 
$ \sigma $ is $ \eps_{\sigma'}/(S-1) $, giving a replication probability from
$ \sigma' $ to $ \sigma $ of $ (\frac{\eps_{\sigma'}}{S-1})^l (1 -
\eps_{\sigma'})^{L - l} $.  Therefore, taking into account the
overall replication rate, we obtain,
\begin{equation}
\kappa_m(\sigma', \sigma) = \kappa_{\sigma'}
(\frac{\eps_{\sigma'}}{S-1})^l (1 - \eps_{\sigma'})^{L - l}
\end{equation}

\section{A Two-Gene Model Incorporating Error Repair}

\subsection{Definitions and Basic Equations}

A simple model to study quasispecies dynamics with genetic repair
is a two gene, single fitness peak (SFP) model.  We take our genome to
have an alphabet size $ S $, composed of ``bases'' $ 0, 1, \dots, S
- 1 $.  The first gene, denoted by $ \sigma_{via} $, has length $
L_{via} $, and controls the viability of the organism.  We assume
that there is a unique, ``fit'' sequence $ \sigma_{via, 0} $ such
that $ \kappa_{\sigma} = k > 1 $ if $ \sigma_{via} = \sigma_{via,
0} $.  Otherwise, $ \kappa_{\sigma} = 1 $.  There is no loss of
generality in assuming $ \kappa_{\sigma} = 1 $ for the unfit
sequences, since time may always be rescaled so that the unfit
$ \kappa_{\sigma} $ become $ 1 $.

The second gene, denoted by $ \sigma_{rep} $, has length $ L_{rep}
$, and is responsible for the enzymatic machinery involved in
repair.  As with viability, there is a unique sequence, $
\sigma_{rep, 0} $, for which repair is functioning, and has a per
base pair failure probability of $ \eps_r $.  For all other $
\sigma_{rep} $ repair is inactivated, and the organism is a
mutator.

For the mutators, the per base pair replication error probability
is taken to be $ \eps $.  Thus, for $ \sigma_{rep, 0} $, the per
base pair replication error probability is $ \eps_r \eps $.  If $
\eps_{\sigma} $ denotes the per base pair replication error
probability of genome $ \sigma $, then $ \eps_{\sigma} = \eps_r
\eps $ if $ \sigma_{rep} = \sigma_{rep, 0} $, and $ \eps $
otherwise.

This model is clearly an oversimplification of the actual genome
and replication dynamics of an organism.  Nevertheless, a two gene, 
SFP model is probably the simplest for studying evolutionary dynamics 
with genetic damage repair, and it is therefore a natural starting point 
before considering more complicated systems.  Despite its simplicity, this 
model still yields sufficiently rich behavior to be of interest.

To determine the equilibrium distribution of genotypes in this
model, note that, by symmetry, we may assume that $
x_{\sigma} $ depends only on the Hamming distance $ l \equiv
HD(\sigma_{via}, \sigma_{via, 0}) $ and $ l' \equiv
HD(\sigma_{rep}, \sigma_{rep, 0}) $.  We define the Hamming class
$ HC(l, l') = \{\sigma = \sigma_{via} \sigma_{rep}|
HD(\sigma_{via}, \sigma_{via, 0}) = l, HD(\sigma_{rep},
\sigma_{rep, 0}) = l'\} $.  Since $ x_{\sigma} $ is assumed to
depend only on the Hamming class of $ \sigma $, we may define $
x_{ll'} = x_{\sigma} $ for $ \sigma \in HC(l, l') $.  We may also
note that $ \kappa_{\sigma} = k $ if $ l = 0 $ and $ 1 $
otherwise, so that $ \kappa_{\sigma} $ depends only on $ l $.
Therefore, we redenote $ \kappa_{\sigma} $ by $ \kappa_l $.
Similarly, we redenote $ \eps_{\sigma} $ by $ \eps_{l'} $.

We wish to express the quasispecies equations in terms of the $
x_{ll'} $.  To do this, we need to sum the mutational
contributions of all $ \sigma $ to the time evolution of $ x_{ll'}
$. Let $ \sigma_{ll'} \in HC(l, l') $.  Any $ \sigma $ may be
obtained from $ \sigma_{ll'} $ by changing the appropriate bases.
Let us write $ \sigma_{ll'} = \sigma_{via, l} \sigma_{rep, l'} $,
and $ \sigma = \sigma_{via} \sigma_{rep} $. By definition of the
Hamming class, $ \sigma_{via, l} $ differs from $ \sigma_{via, 0}
$ in exactly $ l $ places.  Therefore, $ \sigma_{via, l} $ is
identical to $ \sigma_{via, 0} $ in $ L_{via} - l $ places.  Of
these $ L_{via} - l $ bases, let $ l_1 $ denote the number of
bases which are changed in $ \sigma $.  Of the $ l $ bases in $
\sigma_{via, l} $ which are distinct from the corresponding bases
in $ \sigma_{via, 0} $, let $ l_2 $ denote the number which are
changed back to the corresponding bases in $ \sigma_{via, 0} $
when creating $ \sigma $, and let $ l_3 $ denote the number which
are changed to bases which are still distinct from the
corresponding ones in $ \sigma_{via, 0} $.  The base changes
determined by $ l_1 $, $ l_2 $, and $ l_3 $ yields a $
\sigma_{via} $ which is a Hamming distance of $ l_1 + l - l_2 $
from $ \sigma_{via, 0} $, and a Hamming distance of $ l_1 + l_2 +
l_3 $ from $ \sigma_{via, l} $.

For the repair gene, we may define $ l_1' $, $ l_2' $, and $ l_3'
$ similarly.  Thus, given some $ \sigma_{ll'} \in HC(l, l') $, the
vector $ (l_1, l_2, l_3, l_1', l_2', l_3') $ defines a set of base
changes to a $ \sigma_{l_1 + l - l_2, l_1' + l' - l_2'} \in HC(l_1
+ l - l_2, l_1' + l' - l_2') $, such that $ HD(\sigma_{ll'},
\sigma_{l_1 + l - l_2, l_1' + l' - l_2'}) = l_1 + l_2 + l_3 + l_1'
+ l_2' + l_3' $.  We then obtain that,
\begin{widetext}
\begin{equation}
\kappa_m(\sigma_{l_1 + l - l_2, l_1' + l' - l_2'}, \sigma_{ll'}) =
\kappa_{l_1 + l - l_2}(\frac{\epsl}{S-1})^{l_1 +
l_2 + l_3 + l_1' + l_2' + l_3'}(1 - \epsl)^{L_{via} + L_{rep} - l_1 - 
l_2 - l_3 - l_1' - l_2' - l_3'}
\end{equation}
\end{widetext}

The total mutational flow rate into a given $ \sigma_{ll'} $ may
be obtained by summing over the mutational flow rates from all
possible $ (l_1, l_2, l_3, l_1', l_2', l_3') $.  To put together a
final expression, we still need to account for degeneracy, since,
in general, for a given $ \sigma_{ll'} $ and vector $ (l_1, l_2,
l_3, l_1', l_2', l_3') $, there are multiple ways for generating a
new gene sequence.  For a given $ l_1 $, we need to choose $ l_1 $
elements out of $ L_{via} - l $.  Since each selected base can be
changed to $ S - 1 $ other bases, the total number of
possibilities for $ l_1 $ is $ {L_{via} - l \choose
l_1}(S-1)^{l_1} $.  For a given $ l_2 $, we need to choose $ l_2 $
elements out of $ l $.  Since each selected base is restored to
the corresponding base in $ \sigma_{via, 0} $, the total number of
possibilities for $ l_2 $ is $ {l \choose l_2} $. Finally, for a
given $ l_3 $, we need to choose $ l_3 $ elements out of the
remaining $ l - l_2 $. Since each selected base is changed, but is
not changed back to the corresponding element in $ \sigma_{via, 0}
$, there are $ S - 2 $ possibilities per changed base, hence
the total number of possibilities for $ l_3 $ is $ {l - l_2
\choose l_3}(S - 2)^{l_3} $.  Performing a similar analysis for
the repair gene, and putting everything together, we obtain a
total sequence degeneracy of $ {L_{via} - l \choose l_1} {l
\choose l_2} {l - l_2 \choose l_3} {L_{rep} - l' \choose l_1'} {l'
\choose l_2'} {l' - l_2' \choose l_3'} (S-1)^{l_1 + l_1'}
(S-2)^{l_3 + l_3'} $.  Putting everything together, we obtain,
\begin{widetext}
\begin{eqnarray}
\frac{d x_{ll'}}{dt} & = & \sum_{l_1 = 0}^{L_{via} - l} \sum_{l_2
= 0}^{l} \sum_{l_3 = 0}^{l - l_2} \sum_{l_1' = 0}^{L_{rep} - l'}
\sum_{l_2' = 0}^{l'} \sum_{l_3' = 0}^{l' - l_2'}{{L_{via} -
l\choose l_1}{l \choose l_2}{l-l_2 \choose l_3} {L_{rep}-l'
\choose l_1'}{l' \choose l_2'}{l'-l_2' \choose l_3'}
(S-1)^{l_1 + l_1'}(S-2)^{l_3 + l_3'}} \times \nonumber \\
&   & \kappa_{l_1 + l - l_2}(\frac{\epsl}{S-1})^{l_1 + l_2 + l_3 + l_1' 
+ l_2' + l_3'}(1 - \epsl)^{L_{via} + L_{rep} - l_1 - l_2 - l_3 - l_1' - l_2' -
l_3'} x_{l_1 + l - l_2, l_1' + l' - l_2'} - \bar{\kappa}(t)
x_{ll'}
\end{eqnarray}
We may sum over $ l_3 $ and $ l_3' $ to obtain,
\begin{eqnarray}
\frac{d x_{ll'}}{dt} & = & \sum_{l_1 = 0}^{L_{via} - l} \sum_{l_2
= 0} ^{l} \sum_{l_1' = 0}^{L_{rep} - l'} \sum_{l_2' =
0}^{l'}{{L_{via} - l\choose l_1}{l \choose l_2}{{L_{rep} -
l'}\choose{l_1'}}{l' \choose l_2'}\kappa_{l_1 + l -
l_2}(S-1)^{-(l_2 + l_2')}} \times \nonumber \\
&   & \epsl^{l_1 + l_2 + l_1' + l_2'}(1 -
\epsl)^{L_{via} + L_{rep} - l_1 - l_2 - l_1' -
l_2'} (1 + \frac{(S-2)\epsl}{(S-1)(1 -
\epsl)})^{l + l' - l_2 - l_2'} x_{l_1 + l - l_2,
l_1' + l' - l_2'} \nonumber \\
&   & - \bar{\kappa}(t) x_{ll'} \nonumber \\
& = & \sum_{l_1 = 0}^{L_{via} - l} \sum_{l_2 = 0}^{l} \sum_{l_1' =
0}^{L_{rep} - l'} \sum_{l_2' = 0}^{l'}{{L_{via} -
l\choose{l_1}}{l\choose l_2}{L_{rep} - l'\choose l_1'}{l' \choose
l_2'}\kappa_{l_1 + l - l_2}} \times
\nonumber \\
&   & \epsl^{l_1 + l_1'}(1 - \epsl)^{L_{via} + L_{rep} - l - l' - 
l_1 - l_1'}(\frac{\epsl}{S-1})^{l_2 + l_2'}(1 - \frac{\epsl}{S-1})^
{l + l' - l_2 - l_2'} x_{l_1 + l - l_2, l_1' + l' -
l_2'} \nonumber \\
&   & - \bar{\kappa}(t) x_{ll'}
\end{eqnarray}
\end{widetext}
Before proceeding, we introduce the following definitions:
Define $ C_{ll'} $ to be the number of sequences in $ HC(l, l') $.  Note
then that $ C_{ll'} = {L_{via} \choose l}{L_{rep} \choose l'}(S-1)^{l+l'} $.
Define $ z_{ll'} $ to be the fraction of the population in $ HC(l, l') $.
Then $ z_{ll'} = C_{ll'} x_{ll'} $.  Define $ z_0 $ to be the total fraction
of viable organisms, so that $ z_0 = \sum_{l' = 0}^{L_{rep}}{z_{0l'}} $.  
Then $ \bar{\kappa}(t) = (k - 1) z_0 + 1 $.  Reexpressing our dynamical 
equations in terms of the $ z_{ll'} $, we obtain, after some manipulation,
\begin{widetext}
\begin{eqnarray}
\frac{d z_{ll'}}{dt} & = & \sum_{l_1 = 0}^{L_{via} - l} \sum_{l_2
= 0}^{l} \sum_{l_1' = 0}^{L_{rep} - l'} \sum_{l_2' =
0}^{l'}{{L_{via} - l - l_1 + l_2 \choose l_2}{l_1 + l - l_2
\choose l_1}{L_{rep} - l' - l_1' + l_2' \choose l_2'}{l_1' + l' -
l_2' \choose l_1'}}
\kappa_{l_1 + l - l_2} \times \nonumber \\
&   & \epsl^{l_2 + l_2'}(1 - \epsl)^{L_{via} + L_{rep} - l - l' - 
l_1 - l_1'}(\frac{\epsl}{S-1})^{l_1 + l_1'}(1 - \frac{\epsl}{S-1})^
{l + l' - l_2 - l_2'} z_{l_1 + l - l_2, l_1' + l' -
l_2'} \nonumber \\
&   & - ((k - 1) z_0 + 1) z_{ll'}
\end{eqnarray}
\end{widetext}

The equilibrium solution is obtained by solving the equations obtained
by setting the left-hand side to zero.  The numerical solution of the
equilibrium equations is discussed in Appendix A.

\subsection{Behavior in the Limit of Infinite Sequence Length}

We now let the viability and repair sequence lengths $ L_{via}, L_{rep} $
approach $ \infty $, while keeping $ \alpha \equiv L_{via}/L_{rep}
$, $ \mu \equiv L \eps $, and $ \eps_r $ fixed, where $ L \equiv
L_{via} + L_{rep} $ is the total sequence length.  Since $
\epsl = \eps $ or $ \eps_r \eps $, it is clear
that $ \mu_{l_1' + l' - l_2'} \equiv L \epsl $
remains fixed in the limit $ L \rightarrow \infty $.

We claim that, for a given $ l, l' $, the only terms which survive
the limiting process are the $ l_1 = l_1' = 0 $ terms.  We then
note that, as $ L_{via}, L_{rep} \rightarrow \infty $,
\begin{widetext}
\begin{equation}
{L_{via} - l + l_2 \choose l_2} \epsli^{l_2}
\rightarrow \frac{1}{l_2!} (L_{via} \epsli)^{l_2}
= \frac{1}{l_2!}(\frac{\alpha}{\alpha + 1} \mu_{l' -
l_2'})^{l_2}
\end{equation}
and,
\begin{equation}
(1 - \epsli)^{L_{via} - l} \rightarrow
e^{-\frac{\alpha}{\alpha + 1} \mu_{l' - l_2'}}
\end{equation}
Taking similar limits for the $ L_{rep} $ terms, we obtain the
infinite sequence length equations,
\begin{equation}
\frac{d z_{ll'}}{dt} = \sum_{l_2 = 0}^{l} \sum_{l_2' = 0}^{l'}
{\frac{\kappa_{l - l_2}}{l_2! l_2'!} \alpha^{l_2}(\frac{\mu_{l' -
l_2'}}{\alpha + 1})^{l_2 + l_2'} e^{-\mu_{l' - l_2'}} z_{l - l_2,
l' - l_2'}} - \bar{\kappa}(t) z_{ll'}
\end{equation}
Expanding out the terms, and redenoting $ l_2 $ by $ l_1 $, and $
l_2' $ by $ l_1' $, we obtain,
\begin{eqnarray}
\frac{d z_{ll'}}{dt} & = & \frac{k}{l! l'!} \alpha^l (\frac{\eps_r
\mu}{\alpha + 1})^{l + l'} e^{-\eps_r \mu} z_{00} +
\frac{k}{l!}(\frac{\alpha \mu}{\alpha + 1})^l e^{-\mu} \sum_{l_1'
= 0}^{l' - 1}{\frac{1}{l_1'!} (\frac{\mu}{\alpha + 1})^{l_1'}
z_{0, l' - l_1'}} \nonumber \\
&   & + \frac{1}{l'!} (\frac{\eps_r \mu}{\alpha + 1})^{l'}
e^{-\eps_r \mu} \sum_{l_1 = 0}^{l - 1}{\frac{1}{l_1!}(\frac{\alpha
\eps_r \mu}{\alpha + 1})^{l_1} z_{l - l_1, 0}} + e^{-\mu} \sum_{l_1 =
0}^{l - 1} \sum_{l_1' = 0}^{l' - 1}{\frac{1}{l_1!
l_1'!}\alpha^{l_1} (\frac{\mu}{\alpha + 1})^{l_1 + l_1'} z_{l - l_1, l'
- l_1'}} \nonumber \\
&   & - ((k - 1) z_0 + 1) z_{ll'}
\end{eqnarray}
\end{widetext}

To understand why only the $ l_1 = l_1' = 0 $ terms survive, let
us consider the mutational contribution from those $ z_{l_1 + l -
l_2, l_1' + l' - l_2'} $ for which at least one of $ l_1, l_1'
> 0 $.  A $ \sigma' \in HC(l_1 + l - l_2, l_1' + l' - l_2') $ was
obtained from a $ \sigma \in HC(l, l') $ by changing $ l_1 $ of
the $ L_{via} - l $ bases in $ \sigma_{via} $ which were equal to
the corresponding bases in $ \sigma_{via, 0} $, and similarly for
$ l_1' $ and $ \sigma_{rep} $.  Therefore, for $ \sigma'_{via} $
to mutate to $ \sigma_{via} $, $ l_1 $ of the changed bases must
back mutate to the corresponding bases in $ \sigma_{via, 0} $.
However, in the limit of infinite sequence length, the number of
unchanged bases in $ \sigma'_{via} $, given by $ L_{via} - l_1 - l
+ l_2 $, becomes infinite, and so the probability of a mutation
occurring at one of those bases approaches $ 1 $, so that the
probability of back mutation goes to $ 0 $.

This heuristic argument is given a more rigorous justification in
Appendix B.  As a simple check, we also ensure that total
population is conserved in the limiting process.

\section{Solution of the Model}

\subsection{The Phase Diagram}

We begin our solution of the model by computing the equilibrium
values of $ z_0 $ and $ z_{00} $.  We begin with the dynamical
equations for $ z_{0l'} $,
\begin{widetext}
\begin{equation}
\frac{d z_{0l'}}{dt} = \frac{k}{l'!}(\frac{\eps_r \mu}{\alpha +
1})^{l'} e^{-\eps_r \mu} z_{00} + k e^{-\mu} \sum_{l_1' = 0}^{l' -
1}{\frac{1}{l_1'!}(\frac{\mu}{\alpha + 1})^{l_1'} z_{0, l' -
l_1'}} - ((k - 1) z_0 + 1) z_{0l'}
\end{equation}
We may sum from $ l' = 0 - \infty $ to obtain the dynamical
equation for $ z_0 $.  Together with the dynamical equation for $
z_{00} $, we have the pair of equations,
\begin{eqnarray}
\frac{d z_0}{dt} & = & k (e^{-\frac{\alpha}{\alpha + 1} \eps_r
\mu} - e^{-\frac{\alpha}{\alpha + 1} \mu}) z_{00} + (k
e^{-\frac{\alpha}{\alpha + 1} \mu}  - (k-1) z_0 - 1) z_0 \\
\frac{d z_{00}}{dt} & = & (k e^{-\eps_r \mu} - (k - 1)z_0 - 1)
z_{00}
\end{eqnarray}
\end{widetext}
We may obtain the equilibrium solution of these equations by
setting the left hand sides to zero.  A summary of the possible 
solutions is given in Table I.

\begin{table}
\large{
\begin{tabular}{cc}
$ \underline{z_0} $ & $ \underline{z_{00}} $ \\
$ \frac{k e^{-\eps_r \mu} - 1}{k - 1} $ &
$ \frac{e^{-\eps_r \mu} -
 e^{-\frac{\alpha}{\alpha + 1} \mu}}{e^{-\frac{\alpha}{\alpha + 1}
 \eps_r \mu} - e^{-\frac{\alpha}{\alpha + 1} \mu}} z_0 $ \\
$ \frac{k e^{-\frac{\alpha}{\alpha + 1} \mu} - 1}{k - 1} $ & $ 0 $
\\
$ 0 $ & $ 0 $
\end{tabular}
\caption{The possible equilibrium values of $ (z_0, z_{00}) $ as a
function of $ \mu $ and $ \eps_r $.} }
\end{table}

We need to map out the regions in the $ (\mu, \eps_r) $ parameter
space for which the various solutions are valid.  First of all,
note that we must have $ z_0 \in [0, 1] $ and $ z_{00} \in [0,
z_0] $.  For the first solution set to hold, we must therefore
have $ 0 \leq k e^{-\eps_r \mu} - 1 \leq k - 1 $.  The second
inequality is automatically satisfied.  For the first inequality
to hold, we must have $ \eps_r \mu \leq \ln k $.  In order for $
z_{00} \in [0, z_0] $, we must then have $ 0 \leq (e^{-\eps_r \mu}
- e^{-\frac{\alpha}{\alpha+1} \mu})/(e^{-\frac{\alpha}{\alpha+1}
\eps_r \mu} - e^{-\frac{\alpha}{\alpha+1} \mu}) \leq 1 $.  Again,
the second inequality is automatically satisfied, but the first
only holds when $ \eps_r \leq \frac{\alpha}{\alpha + 1} $.
Therefore, the first solution pair is only valid when $ \eps_r \mu
\leq \ln k $, and $ \eps_r \leq \frac{\alpha}{\alpha + 1} $.
However, the other two solution pairs may still yield physical
values for $ (z_0, z_{00}) $ in the domain of validity of the
first solution pair.  To resolve this issue, we note that we
want a solution which gives $ z_{00} \rightarrow 1 $ as $ \eps_r
\rightarrow 0 $.  That is, if repair is perfect, then at
equilibrium the population should only consist of viable
repairers.  Therefore, as $ \eps_r \rightarrow 0 $, we expect the
first solution pair to hold, since it gives the correct limiting
behavior.  By continuity, the first solution pair holds over the
set $ \Omega_1 \equiv \{(\mu, \eps_r) \in [0, \infty) 
\times [0, 1]| \eps_r \mu \leq \ln k, \eps_r \leq \frac{\alpha}
{\alpha + 1} \} $.

As $ \eps_r $ is increased beyond $ \frac{\alpha}{\alpha + 1} $,
the first solution is no longer valid, but the second solution may
still be valid if $ 0 \leq k e^{-\frac{\alpha}{\alpha + 1} \mu} - 1
\leq k - 1 $.  Again, the second inequality is automatically
satisfied, while the first only holds when $ \frac{\alpha}{\alpha
+ 1} \mu \leq \ln k $.  The third solution pair may still be physical
in the domain of validity of the second solution pair. To resolve
this issue, we may note that we want a solution which gives $ z_0
\rightarrow 1 $ as $ \mu \rightarrow 0 $.  That is, in the limit
of no replication errors, all of the population is viable.  Therefore, 
as $ \mu \rightarrow 0 $ with $ \eps_r > \frac{\alpha}{\alpha + 1} $, we 
expect the second solution pair to hold, since it gives the correct limiting 
behavior.  By continuity, the second solution pair holds over the set 
$ \Omega_2 \equiv \{(\mu, \eps_r) \in [0, \infty) \times [0, 1]|
\frac{\alpha}{\alpha + 1} \mu \leq \ln k, \eps_r >
\frac{\alpha}{\alpha + 1} \} $.  The third solution is then the
solution over the domain $ \Omega_3 \equiv ([0, \infty) \times [0,
1])/(\Omega_1 \bigcup \Omega_2) $.

\begin{figure}
\includegraphics[width = 0.7\linewidth]{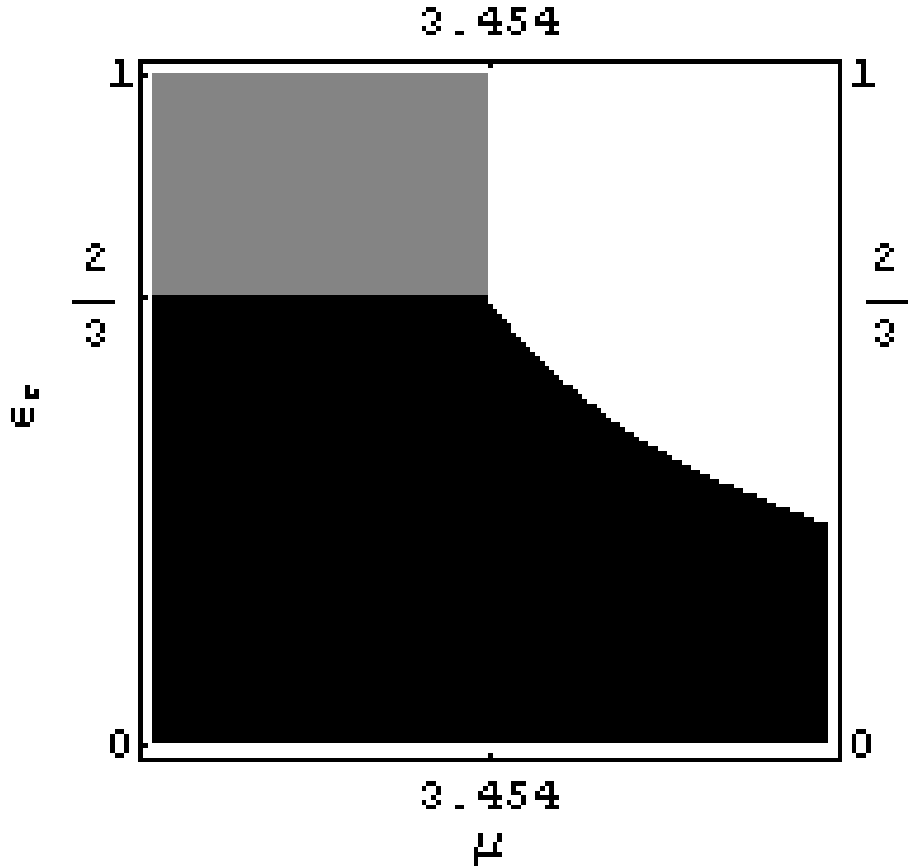}
\caption{Diagram illustrating the solution domains $ \Omega_1 $ (black),
$ \Omega_2 $ (grey), and $ \Omega_3 $ (white).  The $ \mu $ axis
is labelled only at $ \ln k/\eps_{r, crit} \approx 3.454 $, while the 
$ \eps_r $ axis is labelled at $ 0, \eps_{r, crit} = 2/3, 1 $.}
\end{figure}

Figure 1 illustrates the three solution domains $ \Omega_1,
\Omega_2, \Omega_3 $ for $ \alpha = 2, k = 10 $.  In $ \Omega_1 $, the 
population is clustered within finite Hamming distances about the viable and
repair ``master'' sequences.  A finite positive fraction of the
population is viable, and of the viable organisms, a finite
positive fraction of the population is capable of repair.  As $
\eps_r $ is increased beyond $ \frac{\alpha}{\alpha + 1} $, the
population becomes delocalized over the repair gene subspace, and
the fraction of repairers becomes zero.  This phenomenon is known
as the repair catastrophe, and was first predicted in \cite{TDS}.
Nevertheless, if $ \mu $ is still sufficiently small so that $
\frac{\alpha}{\alpha + 1} \mu \leq \ln k $, then the population is
still localized about the viable ``master'' sequence, and the
fraction of viable organisms is positive.  In this regime, as $
\mu $ is increased so that $ \frac{\alpha}{\alpha + 1} \mu > \ln k$, the 
population delocalizes again over the viable gene subspace,
and the fraction of viable organisms drops to zero.  This
phenomenon is known as the error catastrophe.

It is not necessarily true that the repair catastrophe is encountered 
before the error catastrophe.  Defining $ \eps_{r, crit} = 
\frac{\alpha}{\alpha + 1} $, then, whenever $ \eps_r < \eps_{r, crit} $,
the first solution set becomes unphysical when $ \mu > \ln k/\eps_r $.
However, the second solution set is also unphysical, since then $ 
\eps_{r, crit} \mu > \eps_r \mu > \ln k $, so that the third solution
set is the valid one.  Thus, the population goes through the error catastrophe
without going through the repair catastrophe.

The direct transition through the error catastrophe can also occur 
when $ \eps_r $ is varied at fixed $ \mu $.  When $ \mu > \ln k/\eps_{r, crit} 
$, then, as $ \eps_r $ is increased from $ 0 $ to $ 1 $, 
$ k e^{-\eps_r \mu} - 1 $ drops below zero before $ \eps_r = \eps_{r, crit} $.
Therefore, the first solution set becomes unphysical before the repair
catastrophe occurs, but the second solution set is also
unphysical, meaning the third solution set is the one that is
valid.  Thus, for all $ \mu > \ln k/\eps_{r, crit} $, as $ \eps_r
$ is increased from $ 0 $ to $ 1 $, the population undergoes the
error catastrophe at $ \eps_r = \ln k/\mu < \eps_{r, crit} $, so
that the repair catastrophe is never observed. 

\begin{figure}
\includegraphics[width = 0.7\linewidth]{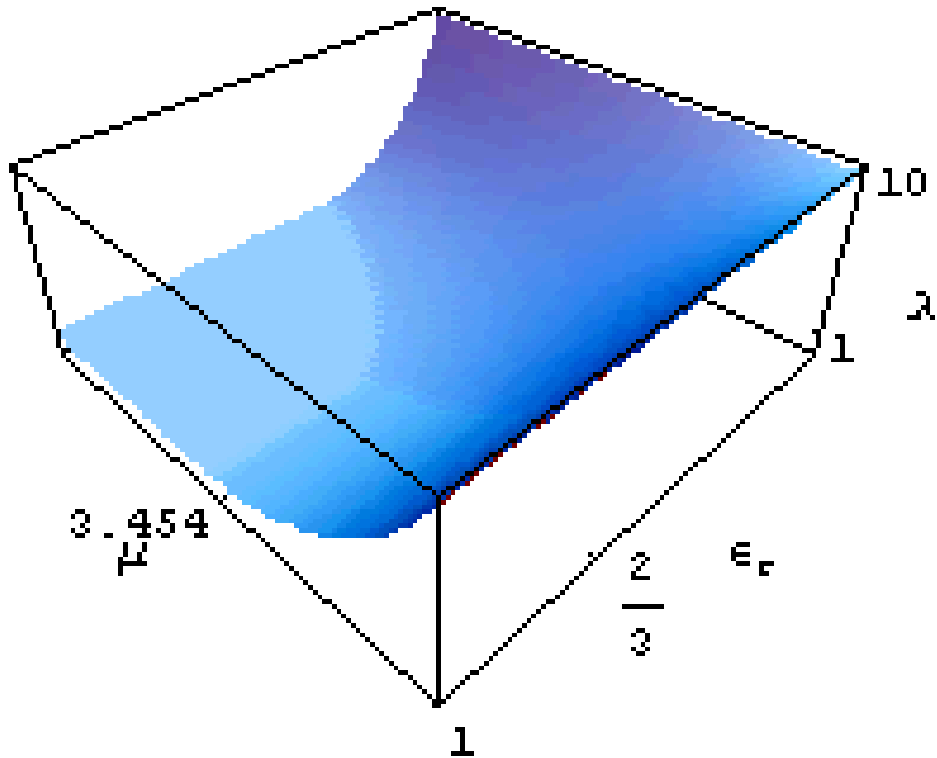}
\caption{Plot of $ \lambda $ for $ \alpha = 2 $, 
$ k = 10 $.}
\end{figure}

We may use our three solution pairs to compute $ \lambda = \bar{\kappa} $
for the three solution domains, or ``phases.''  We have $ \lambda =  
(k - 1) z_0 + 1 $, so that,
\begin{equation}
\lambda(\mu, \eps_r) = \left\{ 
\begin{array}{cc}
k e^{-\eps_r \mu} & \mbox{for } (\mu, \eps_r) \in \Omega_1
\\
k e^{-\frac{\alpha}{\alpha + 1} \mu} & \mbox{for } (\mu, \eps_r) \in \Omega_2 
\\
1 & \mbox{for } (\mu, \eps_r) \in \Omega_3
\end{array}
\right.
\end{equation}

Figure 2 shows a plot of $ \lambda $ versus $ (\mu, \eps_r) $ for 
$ \alpha = 2 $, $ k = 10 $.  Figure 3 shows the corresponding plot for 
$ z_{00} $.  Note that $ \lambda $ is continuous, but not $ \partial 
\lambda/\partial \mu $ and $ \partial \lambda/\partial \eps_r $.  
The error and repair catastrophes are therefore second-order phase 
transitions.  

\begin{figure}
\includegraphics[width = 0.7\linewidth]{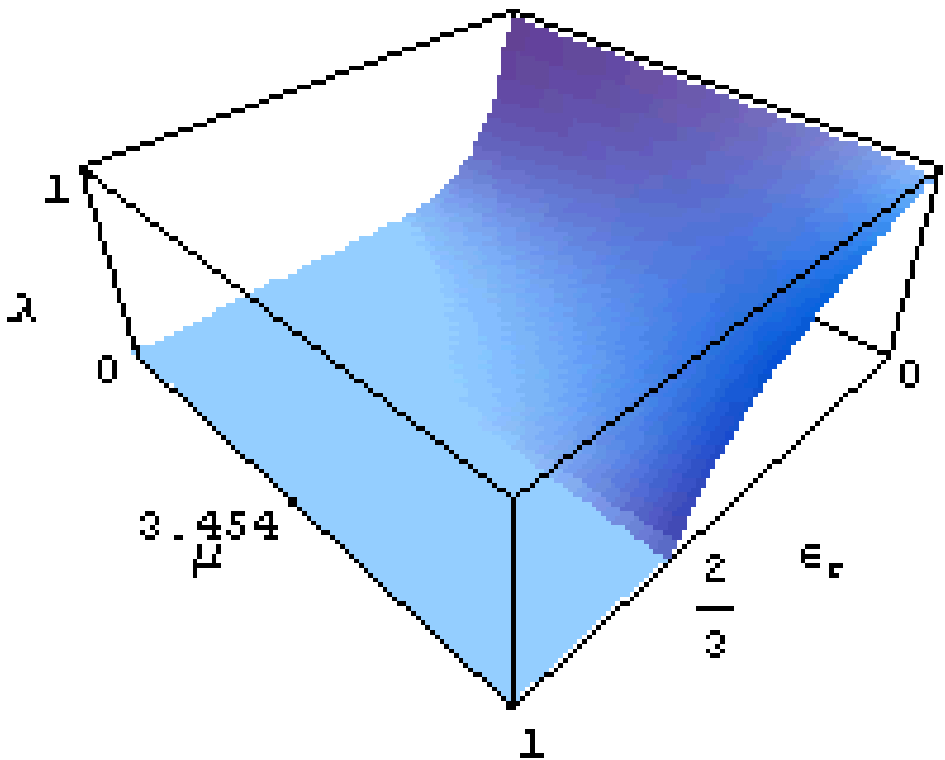}
\caption{Plot of $ z_{00} $ for $ \alpha = 2 $,
$ k = 10 $.}
\end{figure}

The error and repair catastrophes both arise as a result of the interplay
between two competing effects:  (1)  The selective advantage for being viable
and for being a repairer, and (2)  The entropic tendency to be unviable and
a mutator.  For a sufficiently low mutation rate, the selective advantage for
being viable is strong enough to localize the population about 
$ \sigma_{via, 0} $.  However, when the mutation rate exceeds a critical
value, the selective advantage for being viable is no longer sufficiently
strong to localize the population about the viable ``master'' sequence,
and the population delocalizes over the entire viability subspace.  Below
the repair catastrophe, this occurs when the effective growth rate of the
viable, repairing sequence $ \sigma_{via, 0} \sigma_{rep, 0} $ becomes
comparable to the growth rates of the nonviable sequences, i.e., when
$ k e^{-\eps_r \mu} = 1 $.  Above the repair catastrophe, there is no longer
any preference for being a repairer.  The effective growth rate of the
viable sequences due to mutation off of the viability peak is 
$ k e^{-\frac{\alpha \mu}{\alpha + 1}} $, hence, above the repair catastrophe,
the error catastrophe occurs when $ k e^{-\frac{\alpha \mu}{\alpha + 1}} = 1 $.

Below the error catastrophe, viable repairers have a slower rate of mutation
off of the viability peak than viable mutators, and hence have a higher
effective growth rate.  For sufficiently efficient repair, this discrepancy
causes localization about $ \sigma_{rep, 0} $.  However, when the repair
error probability exceeds $ \eps_{r, crit} = \frac{\alpha}{\alpha + 1} = 
L_{via}/L $, the selective advantage for being a repairer is no longer 
sufficient to localize the population, and the distribution undergoes the 
repair catastrophe, in which the distribution delocalizes over the repair 
subspace.  Note that $ \eps_{r, crit} $ is simply the fraction of deleterious
mutations, and increases with increasing $ \alpha $.  This makes sense, since, 
the greater the fraction of deleterious mutations, the greater the relative 
advantage for being a repairer.  Thus, for large $ \alpha $, repair has to be 
highly inefficient before the repair catastrophe occurs.  Conversely, for 
low $ \alpha $, repair has to be highly efficient to give the repairers a 
sufficiently large advantage for the distribution to localize about the 
repair ``master'' sequence.

It should be noted that the error and repair catastrophes are similar to 
thermodynamic phase transitions, in that they both arise from a competition 
between maximum fitness (minimum energy) and maximal entropy.  When the 
replication and repair error probabilities are suficiently low (low 
temperature, high pressure, say), maximal fitness (minimum energy) wins out,
leading to localization on the sequence space.  When the replication or repair 
error probabilities are sufficiently high (high temperature, low pressure), 
maximal entropy wins out, leading to delocalization on the sequence space.  
While not exact, this analogy nevertheless conceptually describes 
the origin of the phases observed in this study.

\subsection{A Recursive Formula for the Population Distribution}

Given $ z_0, z_{00} $, the equilibrium equations may be solved recursively 
to obtain any $ z_{ll'} $ for a given $ (\mu, \eps_r) $ pair.  For $ l' > 0 $,
we have,
\begin{widetext}
\begin{equation}
\frac{d z_{0l'}}{dt} = \frac{k}{l'!}(\frac{\eps_r \mu}{\alpha + 1})^{l'}
e^{-\eps_r \mu} z_{00} + k e^{-\mu} \sum_{l_1' = 1}^{l' - 1}{\frac{1}{l_1'!}
(\frac{\mu}{\alpha + 1})^{l_1'} z_{0, l' - l_1'}} + k e^{-\mu} z_{0l'}
- ((k - 1) z_0 + 1) z_{0l'}
\end{equation}
so at equilbrium we have,
\begin{equation}
z_{0l'} = \frac{1}{(k - 1) z_0 + 1 - k e^{-\mu}}(\frac{k}{l'!}(\frac{\eps_r
\mu}{\alpha + 1})^{l'} e^{-\eps_r \mu} z_{00} + k e^{-\mu} \sum_{l_1' = 1}
^{l' - 1}{\frac{1}{l_1'!} (\frac{\mu}{\alpha + 1})^{l_1'} z_{0, l' - l_1'}})
\end{equation}

We next turn our attention to $ z_{l0} $ for $ l > 0 $.  We have,
\begin{equation}
\frac{d z_{l0}}{dt} = \frac{k}{l!}(\frac{\alpha \eps_r \mu}{\alpha + 1})^{l}
e^{-\eps_r \mu} z_{00} + e^{-\eps_r \mu}\sum_{l_1 = 1}^{l - 1}{\frac{1}{l_1!}
(\frac{\alpha \eps_r \mu}{\alpha + 1})^{l_1} z_{l - l_1, 0}} + 
e^{-\eps_r \mu} z_{l0} - ((k - 1) z_0 + 1) z_{l0}
\end{equation}
so at equilibrium we have,
\begin{equation}
z_{l0} = \frac{1}{(k - 1) z_0 + 1 - e^{-\eps_r \mu}}(\frac{k}{l!}
(\frac{\alpha \eps_r \mu}{\alpha + 1})^{l} e^{-\eps_r \mu} z_{00} + 
e^{-\eps_r \mu}\sum_{l_1 = 1}^{l - 1}{\frac{1}{l_1!}(\frac{\alpha \eps_r \mu}
{\alpha + 1})^{l_1} z_{l - l_1, 0}})
\end{equation}

Finally, we compute the equilibrium value of $ z_{ll'} $ recursively for 
$ l, l' > 0 $.  The result is,
\begin{eqnarray}
z_{ll'} & = & \frac{1}{(k - 1) z_0 + 1 - e^{-\mu}}(\frac{k}{l! l'!} \alpha^l
(\frac{\eps_r \mu}{\alpha + 1})^{l + l'} e^{-\eps_r \mu} z_{00} +
\frac{k}{l!}(\frac{\alpha \mu}{\alpha + 1})^l e^{-\mu} \sum_{l_1' = 0}^{l'-1}
{\frac{1}{l_1'!}(\frac{\mu}{\alpha + 1})^{l_1'} z_{0, l' - l_1'}} 
\nonumber \\ 
&   & + \frac{1}{l'!}(\frac{\eps_r \mu}{\alpha + 1})^{l'} e^{-\eps_r \mu}
\sum_{l_1 = 0}^{l - 1}{\frac{1}{l_1!}(\frac{\alpha \eps_r \mu}
{\alpha + 1})^{l_1} z_{l - l_1, 0}} + e^{-\mu} 
(\sum_{l_1' = 1}^{l' - 1}{\frac{1}{l_1'!} 
(\frac{\mu}{\alpha + 1})^{l_1'} z_{l, l' - l_1'}}
+ \sum_{l_1 = 1}^{l - 1}{\frac{1}{l_1!} (\frac{\alpha \mu}{\alpha + 1})^{l_1}
z_{l - l_1, l'}} \nonumber \\ 
&   & + \sum_{l_1 = 1}^{l - 1}\sum_{l_1' = 1}^{l' - 1}
{\frac{1}{l_1! l_1'!} \alpha^{l_1} (\frac{\mu}{\alpha + 1})^{l_1'}
z_{l - l_1, l' - l_1'}}))
\end{eqnarray}
\end{widetext}

\subsection{Localization Lengths}

The final set of quantities we wish to compute are the following localization 
lengths of the equilibrium distribution:
\begin{eqnarray}
\langle l' \rangle_{via} & \equiv & \sum_{l' = 1}^{\infty}{l' z_{0l'}} \\
\langle l \rangle_{rep} & \equiv & \sum_{l = 1}^{\infty}{l z_{l0}} \\
\langle l \rangle & \equiv & \sum_{l = 1}^{\infty} \sum_{l' = 0}^{\infty}
{l z_{ll'}} \\ 
\langle l' \rangle & \equiv & \sum_{l' = 1}^{\infty} \sum_{l = 0}^{\infty}
{l' z_{ll'}} 
\end{eqnarray}
Using the dynamical equations for the $ z_{ll'} $ we may compute the various
localization lengths at equilibrium.  The basic idea is to obtain an expression
for the time derivatives of the localization lengths in terms of the 
localization lengths themselves, and then solving for the equilibrium value.
We illustrate the technique for $ \langle l' \rangle_{via} $.  We have,
\begin{widetext}
\begin{eqnarray}
\frac{d \langle l' \rangle_{via}}{dt} & = & \sum_{l' = 1}^{\infty}{l' 
\frac{d z_{0l'}}{dt}} \nonumber \\
& = & k \frac{\eps_r \mu}{\alpha + 1} e^{-\frac{\alpha}{\alpha + 1} \eps_r \mu}
z_{00} + k e^{-\mu}
\sum_{l_1' = 0}^{\infty}\sum_{l' = l_1' + 1}^{\infty}{\frac{l' - l_1' + l_1'}
{l_1'!}(\frac{\mu}{\alpha + 1})^{l_1'} z_{0, l' - l_1'}} - ((k - 1) z_0 + 1)
\langle l' \rangle_{via} \nonumber \\
& = & k \frac{\eps_r \mu}{\alpha + 1} e^{-\frac{\alpha}{\alpha + 1} \eps_r \mu}
z_{00} + k e^{-\mu}
\sum_{l_1' = 0}^{\infty}{\frac{1}{l_1'!}(\frac{\mu}{\alpha + 1})^{l_1'}
(\langle l' \rangle_{via} + l_1' (z_0 - z_{00}))} - ((k - 1) z_0 + 1)
\langle l' \rangle_{via} \nonumber \\
& = & k \frac{\eps_r \mu}{\alpha + 1} e^{-\frac{\alpha}{\alpha + 1} \eps_r \mu}
z_{00} + k e^{
-\frac{\alpha}{\alpha + 1} \mu} (\langle l' \rangle_{via} + \frac{\mu}
{\alpha + 1} (z_0 - z_{00})) - ((k - 1) z_0 + 1) \langle l' \rangle_{via}
\end{eqnarray}
so at equilibrium we obtain,
\begin{equation}
\langle l' \rangle_{via} = \frac{k \mu}{\alpha + 1}
\frac{(\eps_r e^{-\frac{\alpha}{\alpha +1} \eps_r \mu}
- e^{-\frac{\alpha}{\alpha + 1} \mu}) z_{00} +
e^{-\frac{\alpha}{\alpha + 1} \mu} z_0}{(k - 1) z_0
+ 1 - k e^{-\frac{\alpha}{\alpha + 1} \mu}}
\end{equation}
To compute the remaining localization lengths using the above approach,
we first need to compute $ z_0' \equiv \sum_{l = 0}^{\infty}{z_{l0}} $.  
Note that $ z_0' $ is simply the total fraction of repairers.  We compute
$ z_0' $ by evaluating $ d z_0'/dt = \sum_{l = 0}^{\infty}
{d z_{l0}/dt} $.  The result is an expression in terms of $ z_0', z_0 $,
and $ z_{00} $, which may be solved at equilibrium to obtain,
\begin{equation}
z_0' = \frac{(k - 1) e^{-\frac{\eps_r \mu}{\alpha + 1}}}{(k - 1) z_0 +
1 - e^{-\frac{\eps_r \mu}{\alpha + 1}}} z_{00}
\end{equation}
We then obtain,
\begin{eqnarray}
\langle l \rangle_{rep} & = & \frac{\alpha \eps_r \mu}{\alpha + 1} 
e^{-\frac{\eps_r \mu}{\alpha + 1}}\frac{(k - 1) z_{00} + z_0'}
{(k - 1) z_0 + 1 - e^{-\frac{\eps_r \mu}{\alpha + 1}}} \\
(\langle l \rangle, \langle l' \rangle) & = & \frac{1}{z_0}(0, \langle
l' \rangle_{via}) + \frac{\mu}{(k - 1)(\alpha + 1) z_0}
((k - 1) z_0 + 1 - (1 - \eps_r) z_0' - (k - 1)(1 - \eps_r) z_{00})(\alpha, 1)
\end{eqnarray}
\end{widetext}

\subsection{Limiting Forms of the Distribution}

It is instructive to study the behavior of the distribution in the
following limiting cases:  (1) $ \mu \rightarrow 0 $.  (2) $ \alpha 
\rightarrow \infty $.  (3) $ \alpha \rightarrow 0 $.  We handle each of these 
cases in turn.

\subsubsection{Behavior in the Limit $ \mu \rightarrow 0 $}

For $ \mu \rightarrow 0 $, note that $ z_0 \rightarrow 1 $,
and $ z_{00} \rightarrow (\frac{\alpha}{\alpha + 1} - \eps_r)/(\frac{\alpha}
{\alpha + 1}(1 - \eps_r)) = 1 - \eps_r/(\alpha (1 - \eps_r)) $, below the
repair catastrophe.  We may also note that $ z_0' \rightarrow z_{00} $, and 
$ (\langle l \rangle, \langle l' \rangle) \rightarrow (0, \langle l' 
\rangle_{via}) $.  This makes sense since in the limit $ \mu \rightarrow 0 $, 
we expect that the entire population becomes viable.  For the same reason, 
$ \langle l \rangle_{rep} \rightarrow 0 $ as $ \mu \rightarrow 0 $.  Finally, 
as $ \mu \rightarrow 0 $ for $ \eps_r < \eps_{r, crit} $, $ \langle l' 
\rangle_{via} \rightarrow \frac{1}{(\alpha + 1)(\frac{\alpha}{\alpha+1} - 
\eps_r)}(1 - (1 - \eps_r)(1 - \frac{\eps_r}{\alpha (1 - \eps_r)})) = 
\frac{1}{(\alpha + 1)(\eps_{r, crit} - \eps_r)}\eps_r \frac{\alpha + 1}
{\alpha} = (1 - \eps_{r, crit})/\eps_{r, crit} \times \eps_r/(\eps_{r, crit} -
\eps_r) $.  As expected, these results agree with the point-mutation limit 
expressions obtained in \cite{TDS}.

\subsubsection{Behavior in the Limit $ \alpha \rightarrow \infty $}

For $ \alpha \rightarrow \infty $ we obtain $ \eps_{r, crit} = 1 $.  Hence,
we are always below the repair catastrophe.  As long as $ \mu <
\ln k/\eps_r $, then $ z_{00} \rightarrow z_0 = (k e^{-\eps_r \mu} - 1)/(k - 1)
$.  Thus, the solution pairs presented in Table I reduce to two possible
solutions.  Either $ z_0 = z_{00} = (k e^{-\eps_r \mu} - 1)/(k - 1) $ 
if we are below the error catastrophe ($ \eps_r \mu < \ln k $), 
or $ z_0 = z_{00} = 0 $ if we are above the error catastrophe.  This means 
that the fraction of mutators is always zero.  To understand this behavior, 
note that the probability of mutating off of the repairer sequence is 
$ 1 - e^{-\frac{\eps_r \mu}{\alpha + 1}} $, while the probability of
mutating off of a mutator sequence is $ 1 - e^{-\frac{\mu}{\alpha + 1}}
$.  Both go to $ 0 $ as $ \alpha \rightarrow \infty $.  However, since for
$ \eps_r < 1 $ the repairer sequence has a greater selective advantage than
the mutator sequence, the repair strain comes to dominate the population.  
Only at $ \eps_r = 1 $ is there an ambiguity, since $ (e^{-\eps_r \mu} - 
e^{-\frac{\alpha \mu}{\alpha + 1}})/(e^{-\frac{\alpha \eps_r \mu}{\alpha + 1}}
- e^{-\frac{\alpha \mu}{\alpha + 1}}) \rightarrow 0/0 $, which is undefined.
Physically, since at $ \eps_r = 1 $ there is no difference between what we
call a ``repairer'' and a ``mutator,'' we expect delocalization over the
repair subspace, so that $ z_{00} \rightarrow 0 $.

We may also note that $ z_0' \rightarrow z_{00}/z_0 = 1 $ for $ \eps_r < 1 $,
and $ 0 $ for $ \eps_r = 1 $.  We also have $ \langle l \rangle_{rep} 
\rightarrow k \eps_r \mu e^{-\eps_r \mu}/(k e^{-\eps_r \mu} - 1) $.  Also, 
$ \langle l' \rangle_{via} \rightarrow 0 $, for $ \eps_r < 1 $, and 
$ \infty $ for $ \eps_r = 1 $.

\subsubsection{Behavior in the Limit $ \alpha \rightarrow 0 $}

For $ \alpha \rightarrow 0 $, we have $ \eps_{r, crit} \rightarrow 0 $.
Therefore, for all $ \eps_r > 0 $ we are beyond the repair catastrophe,
and since $ \frac{\alpha}{\alpha + 1} \mu = 0 < \ln k $, we are below
the error catastrophe as well, so that $ z_0 = 1 $ with $ z_{00} = 0 $. 
This makes sense, since, for $ \alpha = 0 $, the probability of mutating
off of the viability peak is $ 1 - e^{-\frac{\alpha}{\alpha + 1} \mu} 
\rightarrow 0 $.  Thus, the entire population is viable at equilibrium.
As for $ z_{00} $, we note that $ z_{00} = 0 $ for $ \eps_r > 0 $, but for
$ \eps_r = 0 $ we obtain the expression, $ (e^{0} - e^{0})/(e^{0} - e^{0})
\times z_0 = 0/0 $.  Physically, we must have $ z_{00} = 1 $ at $ \eps_r
= 0 $.  This ambiguity is therefore resolved by letting $ \alpha \rightarrow
0 $ for $ \eps_r = 0 $.  That is, we evaluate $ z_{00} $ for $ \alpha = 0,
\eps_r = 0 $ by setting $ z_{00}\lvb_{\alpha = 0, \eps_r = 0} =
\lim_{\alpha \rightarrow 0}{z_{00}\lvb_{\alpha, \eps_r = 0}} $.

As expected, for $ \eps_r > 0 $ we have $ z_0' = 0 $, $ \langle l' 
\rangle_{via} = \infty $, $ \langle l \rangle_{rep} = 0 $, 
$ \langle l \rangle = 0 $, and $ \langle l' \rangle = \infty $.  Again,
for $ \eps_r = 0 $ we resolve any ambiguities by letting
$ \alpha \rightarrow 0 $, giving, as expected, $ z_0' = 1 $, 
$ \langle l' \rangle_{via} = \langle l \rangle_{rep} = \langle l \rangle = 
\langle l' \rangle = 0 $.

\section{Conclusions and Future Research}

This paper presented a two gene, single fitness peak model to determine the 
equilibrium distribution of genotypes in a unicellular population capable of 
replication error repair.  The work presented here was a continuation of 
\cite{TDS}, in which the equilibrium distribution of mutators was studied for 
mutation rates well below the error catastrophe.  This paper obtained the 
equilibrium behavior of the two gene model for arbitrary mutation rates, 
thereby incorporating both the error and repair catastrophes into a single, 
two-dimensional phase diagram.  While our model is probably the simplest one 
could use for studying evolutionary dynamics in the presence of genetic repair,
it does nevertheless make experimentally testable predictions.  As mentioned 
in the Introduction, the error catastrophe has already been observed 
\cite{CAT1, CAT2}.  The repair catastrophe would be more difficult to observe 
experimentally, since it would be necessary to selectively interfere with the 
DNA mismatch repair system.  If possible, however, it would be interesting to 
try to experimentally map out the phase diagram shown in Figure 1 for an 
actual organism, such as {\it E. coli}.

In \cite{TDS} it was noted that the equilibrium distribution of mutators did 
not depend on $ \mu $, but only on $ \eps_r $ and $ \alpha $.  This was 
interesting since the larger the value of $ \mu $, the greater the difference 
in mutation rates off of the viability peak between repairers and mutators.
One might also naively expect the repair catastrophe to disappear entirely as
$ \mu \rightarrow 0 $, since the difference in viability between repairers
and mutators disappears in the limit of no mutations.  In \cite{TDS} 
it was assumed that mutations were sufficiently slow so that only point
mutations needed to be considered.  In the complete model, when we 
allow for mutations between any two genomes, we do indeed obtain a 
$ \mu $-dependence on the equilibrium distribution of mutators.  Interestingly,
however, the repair catastrophe still occurs at $ \eps_{r, crit} =
\frac{\alpha}{\alpha + 1} $, unchanged from the point-mutation result in
\cite{TDS}.

We believe that the solution technique developed in this paper may be used
to solve a large class of mutation dynamics equations.  To illustrate,
consider a more general genome consisting of $ N $ genes, so that the full 
sequence $ \sigma $ may written as $ \sigma = \sigma_1 \dots \sigma_N $.
We assume that there exist ``master'' sequences $ \sigma_{1,0}, \dots,
\sigma_{N,0} $, such that the properties of each $ \sigma_n $ depends only
on the Hamming distance $ HD(\sigma_n, \sigma_{n, 0}) $.  We then can define
the Hamming class $ HC(l_1, \dots, l_N) = \{\sigma = \sigma_1 \dots \sigma_N
| HD(\sigma_n, \sigma_{n,0}) = l_n, n = 1, \dots, N\} $.  Consider then
some $ \sigma_{l_1, \dots, l_N} \in HC(l_1, \dots, l_N) $.  For each
$ n \in \{1, \dots, N\} $, define $ l_{n1}, l_{n2}, l_{n3} $ analogously
to $ l_1, l_2, l_3 $ and $ l_1', l_2', l_3' $ from Section III.A.  Then the
vector $ (l_{11}, l_{12}, l_{13}, \dots, l_{N1}, l_{N2}, l_{N3}) $ defines
a set of base changes to a new genome $ \sigma_{l_{11} + l_1 - l_{12},
\dots, l_{N1} + l_N - l_{N2}} \in HC(l_{11} + l_1 - l_{12}, \dots,
l_{N1} + l_N - l_{N2}) $, which is a Hamming distance of $ l_{11} + l_{12}
+ l_{13} + \dots + l_{N1} + l_{N2} + l_{N3} $ from $ \sigma_{l_1, \dots,
l_N} $.  Proceeding as in Section III.A, we may define $ z_{l_1, \dots, l_N} $
to be the fraction of the population in $ HC(l_1, \dots, l_N) $, and obtain an 
expression for $ d z_{l_1, \dots, l_N}/dt $ which is a generalization of
the expression given in Eq. (9).  Presumably, the back mutation terms should
drop out in the limit of infinite sequence length, giving an infinite 
sequence length equation similar to Eq. (12).

For future research, we would like to move away from studies of equilibria
and focus on the role that mutators play in dynamic environments.  
Incorporating the effects of horizontal transfer between organisms will
also be useful for exploring phenomenological aspects of antibiotic drug
resistance.  We also seek to develop more realistic replication
models, incorporating the double-stranded nature of the DNA molecule.
In our current model, we essentially ``black-boxed'' the replication dynamics,
and assumed that the double-stranded DNA could be represented as a single
symbol sequence.  While the complementary nature of the double helix makes
this assumption technically correct, the actual replication dynamics with 
two complementary strands is somewhat different than the single-strand model
used in this paper.  

Finally, we plan to develop collaborations with experimental groups
working on mismatch repair, and attempt to devise possible strategies for
tuning the efficiency of the mismatch repair system.  If successful, such
experiments would give direct confirmation of the repair catastrophe, and 
provide a better understanding of error correction mechanisms in 
biological systems.

\begin{acknowledgments}

This research was supported by an NIH postdoctoral research
fellowship.

\end{acknowledgments}

\begin{appendix}

\section{Numerical Solution of the Model for Finite Genomes}

Equation (9) in Section III.A gives the expression for the Hamming class 
symmetrized dynamics equations of our model.  We can put this equation into 
matrix form,
\begin{equation}
\frac{d \vec{z}}{dt} = B \vec{z} - (\vec{\kappa} \cdot \vec{z}) \vec{z}
\end{equation}
where $ \vec{z} = (z_{ll'}) $ is the vector of population fractions in the 
various Hamming classes, $ B $ is the matrix of first-order mutation rate
constants between the various Hamming classes, and $ \vec{\kappa} $ is
the vector of first-order growth rate constants for the various Hamming
classes, so that $ \vec{\kappa} \cdot \vec{z} = \sum_{l = 0}^{L_{via}}
\sum_{l' = 0}^{L_{rep}}{\kappa_{ll'} z_{ll'}} $, where $ \kappa_{ll'} $
is simply $ \kappa_{l} $ in our model.

The equilibrium distribution may then be solved using fixed-point iteration,
via the equation,
\begin{equation}
\vec{z}_{n+1} = \frac{1}{\vec{\kappa} \cdot \vec{z_{n}}} B \vec{z_n}
\end{equation}
In principle, the iterations are terminated when the $ z_n $ stop
changing.  We introduce a fractional cutoff parameter $ \delta $,
and stop iterating when $ (z_{n+N_\eps, ll'} - z_{n, ll'})/z_{n, ll'}  
< \delta $.  $ N_\eps $ is chosen to be sufficiently large so that on the 
order of one mutation is allowed to occur after $ N_\eps $ iterations, to 
ensure that equilibration is being accurately measured.  For a large sequence 
length $ L $, the probability of correct replication is $ e^{-L \eps} $, so the
probability of incorrect replication is $ 1 - e^{-L \eps} $.  Therefore,
taking $ N_\eps = 1/(1 - e^{-L \eps}) $ ensures that on the order of one
incorrect replication has occurred, so that if $ (z_{n + N_\eps, ll'} - 
z_{n, ll'})/z_{n, ll'} < \delta $ for all $ l, l' $, then it is
possible to assume that equilibration has been achieved.

Note that what this method does is account for the fact that equilibration
takes longer for smaller values of $ \eps $, i.e., for slower mutation
rates.  Since $ \lim_{\eps \rightarrow 0} {N_\eps} = \infty $, and
$ \lim_{\eps \rightarrow 1} {N_\eps} \approx 1 $ for large $ L $, we see that 
our choice of $ N_\eps $ accounts for the slower equilibration rate by 
iterating more times before comparing the changes in the $ z_{ll'} $.  In our 
numerical simulations, we found that $ \delta = 10^{-4} - 10^{-3} $ was 
sufficient to achieve good convergence.  For $ \alpha = 2 $, $ k = 10 $, it 
was found that for $ L = 30 $ the equilibrium values of $ z_0 $ and $ z_{00} $ 
were almost identical to their $ L = \infty $ values.  For this reason, we did
not give figures showing the results of numerical simulations in this paper.

\section{Justification of the Infinite Sequence Length Form of the Dynamical 
Equations}

To establish the infinite sequence length form of Eq. (9) in Section
III.A, we need to first establish some basic inequalities, to facilitate 
the computation of upper bounds.  We begin with the following inequality, 
for $ l_1 > 0 $:
\begin{widetext}
\begin{equation}
{l_1 + l - l_2 \choose l_1} (\frac{\epsl}
{S - 1})^{l_1} (1 - \frac{\epsl}{S-1})^{l - l_2} \leq
(\frac{\eps}{S-1})^{l_1} \prod_{k = 1}^{l_1}{\frac{k + l - l_2}{k}}  = 
(\frac{\eps}{S-1})^{l_1} \prod_{k = 1}^{l_1}{(1 + \frac{l - l_2}{k})} \leq 
(\frac{l+1}{S-1} \eps)^{l_1}
\end{equation}
Note that this inequality also holds for $ l_1 = 0 $.  A similar inequality 
holds for the primed indices.  Our next inequality is
simply,
\begin{equation}
{L_{via} - l - l_1 + l_2 \choose l_2} \epsl^{l_2}
(1 - \epsl)^{L_{via} - l - l_1} \leq 1
\end{equation}
and similarly for the primed indices.  Finally, we may note that $ z_{ll'}
\leq 1 $ for all $ l, l' $.  Now, to simplify the calculation, denote the 
summand in Eq. (9) of Section III.A by $ S_{l l_1 l_2 l' l_1' l_2'} $.  Then 
putting together our inequalities, we obtain,
\begin{eqnarray}
\sum_{l_2 = 0}^{l}\sum_{l_2' = 0}^{l'}{S_{l 0 l_2 l' 0 l_2'}} 
& \leq &
\sum_{l_1 = 0}^{L_{via} - l}\sum_{l_2 = 0}^{l}\sum_{l_1' = 0}^{L_{rep} - l'}
\sum_{l_2' = 0}^{l'}{S_{l l_1 l_2 l' l_1' l_2'}} \nonumber \\
& \leq &
\sum_{l_2 = 0}^{l}\sum_{l_2' = 0}^{l'}{S_{l 0 l_2 l' 0 l_2'}} +
\sum_{l_1 = 1}^{L_{via} - l}\sum_{l_2 = 0}^{l}\sum_{l_2' = 0}^{l'}
{k (\frac{l+1}{S-1} \eps)^{l_1}} \nonumber \\
&   & + \sum_{l_1' = 1}^{L_{rep} - l'}\sum_{l_2' = 0}^{l'}\sum_{l_2 = 0}^{l}
{k (\frac{l'+1}{S-1} \eps)^{l_1'}} + 
\sum_{l_1 = 1}^{L_{via} - l}\sum_{l_2 = 0}^{l}\sum_{l_1' = 1}^
{L_{rep} - l'}\sum_{l_2' = 0}^{l'}{k (\frac{l+1}{S-1} \eps)^{l_1} 
(\frac{l'+1}{S-1} \eps)^{l_1'}} \nonumber \\
& = & 
\sum_{l_2 = 0}^{l}\sum_{l_2' = 0}^{l'}{S_{l 0 l_2 l' 0 l_2'}} +
k (l+1)^2 (l'+1) \frac{\eps}{S-1} \frac{1 - (\frac{l+1}{S-1}\eps)^
{L_{via} - l}}{1 - \frac{l+1}{S-1}\eps} \nonumber \\
&   & + k (l+1) (l'+1)^2 \frac{\eps}{S-1} \frac{1 - (\frac{l'+1}{S-1}\eps)^
{L_{rep} - l'}}{1 - \frac{l'+1}{S-1}\eps} \nonumber \\
&   & + k (l+1)^2 (l'+1)^2 (\frac{\eps}{S-1})^2 \frac{1 - 
(\frac{l+1}{S-1}\eps)^{L_{via} - l}}{1 - \frac{l+1}{S-1}\eps} 
\frac{1 - (\frac{l'+1}{S-1}\eps)^{L_{rep} - l'}}{1 - \frac{l'+1}{S-1}\eps}
\end{eqnarray}
Now, following the argument from the beginning of Section III.B, we have
that, as $ L_{via}, L_{rep} \rightarrow \infty $ at fixed $ \alpha, \mu,
\eps_r $, we get that,
\begin{equation}
\sum_{l_2 = 0}^{l}\sum_{l_2' = 0}^{l'}{S_{l 0 l_2 l' 0 l_2'}} 
\rightarrow \sum_{l_2 = 0}^{l}\sum_{l_2' = 0}^{l'}
{\frac{\kappa_{l - l_2}}{l_2! l_2'!} \alpha^{l_2} (\frac{\mu_{l' - l_2'}}
{\alpha + 1})^{l_2 + l_2'} e^{-\mu_{l' - l_2'}} z_{l - l_2, l' - l_2'}}
\end{equation}
hence, since $ \eps \rightarrow 0 $ at fixed $ \mu $ when $ L_{via}, L_{rep}
\rightarrow \infty $, we see from the inequalities given in Eq. (B3) that,
\begin{equation}
\sum_{l_1 = 0}^{L_{via} - l}\sum_{l_2 = 0}^{l}\sum_{l_1' = 0}^{L_{rep} - l'}
\sum_{l_2' = 0}^{l'}{S_{l l_1 l_2 l' l_1' l_2'}} \rightarrow
\sum_{l_2 = 0}^{l}\sum_{l_2' = 0}^{l'}
{\frac{\kappa_{l - l_2}}{l_2! l_2'!} \alpha^{l_2} (\frac{\mu_{l' - l_2'}}
{\alpha + 1})^{l_2 + l_2'} e^{-\mu_{l' - l_2'}} z_{l - l_2, l' - l_2'}}
\end{equation}
The convergence is not uniform, however, since our upper bound depends on
$ l, l' $.  

This establishes the infinite sequence length form of our dynamical equations,
as given in Eq. (12) of Section III.B.  We may verify that total probability
is conserved in our limiting process.  Defining $ z = \sum_{l,l'}{z_{ll'}} $,
we obtain,
\begin{eqnarray}
\frac{d z}{dt} & = & \sum_{l = 0}^{\infty} \sum_{l' = 0}^{\infty}
\sum_{l_1 = 0}^{l} \sum_{l_1' = 0}^{l'}{\frac{\kappa_{l - l_1}}{l_1! l_1'!}
\alpha^{l_1} (\frac{\mu_{l' - l_1'}}{\alpha + 1})^{l_1 + l_1'}
e^{-\mu_{l' - l_1'}} z_{l - l_1, l' - l_1'}} - \bar{\kappa}(t) z \nonumber \\
& = & \sum_{l_1 = 0}^{\infty}\sum_{l_1' = 0}^{\infty}\sum_{l = l_1}^{\infty}
\sum_{l' = l_1'}^{\infty}{\frac{\kappa_{l - l_1}}{l_1! l_1'!}
\alpha^{l_1} (\frac{\mu_{l' - l_1'}}{\alpha + 1})^{l_1 + l_1'}
e^{-\mu_{l' - l_1'}} z_{l - l_1, l' - l_1'}} - \bar{\kappa}(t) z \nonumber \\
& = & \sum_{k_1 = 0}^{\infty}\sum_{k_1' = 0}^{\infty}{
\kappa_{k_1} e^{-\mu_{k_1'}} z_{k_1, k_1'} \sum_{l_1 = 0}^{\infty}
\sum_{l_1' = 0}^{\infty}{\frac{1}{l_1! l_1'!} \alpha^{l_1} 
(\frac{\mu_{k_1'}}{\alpha + 1})^{l_1 + l_1'}}} - \bar{\kappa}(t) z \nonumber \\
& = & \sum_{k_1 = 0}^{\infty}\sum_{k_1' = 0}^{\infty}
{\kappa_{k_1} z_{k_1, k_1'}} - \bar{\kappa}(t) z \nonumber \\
& = & \bar{\kappa}(t) z - \bar{\kappa}(t) z = 0
\end{eqnarray} 
Thus, since $ z $ starts off at $ 1 $, it remains $ 1 $ at all times, hence
total probability is conserved in the infinite sequence limit.

\end{widetext}

\end{appendix}

\bibliography{mu_catas_full}

\end{document}